\documentclass[aps,prl,reprint,article,floatfix]{revtex4-1}
\pdfoutput=1 

\usepackage{fullpage}
\usepackage{amsmath}
\usepackage{amssymb}
\usepackage{graphicx}

\begin{document}

\title{Critical Black Holes in a Large Charge Limit}
\author{Clifford V. Johnson}
\email{johnson1@usc.edu}
\affiliation{Department of Physics and Astronomy\\ University of
Southern California \\
 Los Angeles, CA 90089-0484, U.S.A.}

\pacs{05.70.Ce,05.70.Fh,04.70.Dy}

\begin{abstract}
The extended thermodynamics of  Reissner--Nordstr\"om charged black holes in $D$--dimensional anti--de Sitter spacetime has a phase structure in the $(T,p)$  plane that includes a line of first order phase transitions ending in a second order transition point. For each $D\geq4$, properties of the critical black hole at that point are explored, motivated by the observation that in $D=4$ its neighbourhood plays a core role in defining a very special class  of heat engines that asymptote to Carnot efficiency in the large charge limit. This new limit provides a model of a novel low--temperature thermodynamic system. The double  limit of  approaching the horizon while sending $q$ large at an appropriate rate yields a fully decoupled  $D$--dimensional Rindler spacetime with zero cosmological constant.

\end{abstract}


\maketitle

The thermodynamics arising from the semi--classical quantum treatment of gravitational systems (sometimes involving, but not limited to, black hole spacetimes \cite{Bekenstein:1973ur,Bekenstein:1974ax,Hawking:1974sw,Hawking:1976de,Gibbons:1976ue}) define interesting equations of state that are of interest beyond the  gravitational context. When the spacetime metric is known, simple geometric techniques can supply explicit equations for many thermodynamic quantities that would have been hard to compute using more traditional approaches. This is especially true now that the extended thermodynamics\cite{Kastor:2009wy} supplies a pressure $p$ and volume $V$ to supplement the vocabulary\footnote{See ref. \cite{Kubiznak:2016qmn} for a review of some of the  results obtained  in this context by using the extended thermodynamics framework.}.

This Letter explores properties of a class of charged black holes in asymptotically anti--de Sitter spacetimes, focussing on a special point in parameter space where they lie at a critical point, as will be reviewed below.  Of particular interest will be a new limit where, staying at the critical point, the black hole charge will be taken to be large. It is motivated by recent work \cite{Johnson:2017hxu} in which a certain kind of heat engine (that uses the black hole's equation of state as a definition of the working substance \cite{Johnson:2014yja}) working close to the critical regime, is driven to Carnot efficiency at large charge. The properties of the critical black holes themselves merit further exploration. In what follows, after some reminders of the key features of the context, and of the motivation, some interesting new results for them are uncovered and discussed.

The context will be a $D$--dimensional  Einstein--Maxwell system, for $D\geq4$, with action\footnote{Here, the conventions of refs.\cite{Chamblin:1999tk,Chamblin:1999hg} will be used.}:
\begin{equation}
I=-\frac{1}{16\pi }\int \! d^Dx \sqrt{-g} \left(R-2\Lambda -F^2\right)\ .
\label{eq:action}
\end{equation}
  The cosmological constant $\Lambda=-(D-1)(D-2)/2l^2$  sets a length scale~$l$, and we have set Newton's constant $G$ and the speed of light $c$ to unity (as we will later for $\hbar$ and $k_B$). The black hole  spacetimes of interest here are Reissner--Nordstr\"om--like solutions.  The metric is:
\begin{eqnarray}
ds^2 &=& -Y( r)dt^2
+ {dr^2\over Y(r)} + r^2 d\Omega_{D-2}^2\ , \\
Y( r) &\equiv& 1-\frac{m}{r^{D-3}}+\frac{q^2}{r^{2D-6}}+\frac{r^2}{l^2}\ , 
\end{eqnarray}
where $d\Omega_{D-2}$ is the metric on a round $S^{D-2}$, $t$ is time and $r$ is a radial coordinate. The gauge potential is:
\begin{eqnarray}
 A_t = \frac{q}{c}\left(\frac{1}{r_+^{D-3}}-\frac{1}{r^{D-3}}\right)\ , \,\,\, c=\sqrt{\frac{2(D-3)}{(D-2)}} \ .
\end{eqnarray}
  The parameters $m$ and $q$ set a mass and charge according to:
\begin{eqnarray}
\label{eq:mass}
M&=&\frac{(D-2)\omega_{D-2}}{16\pi}m\ ,\,\,\,{\rm and}\\
\label{eq:charge}
Q&=&\sqrt{2(D-2)(D-3)}\left(\frac{\omega_{D-2}}{8\pi}\right)q\ ,
\end{eqnarray}
where $\omega_{D-2}$ is the volume of the round $S^{D-2}$ surface. The potential  has been chosen to vanish on the outer horizon  at $r=r_+$, the largest positive real root of $Y(r)$.

 Several aspects of the thermodynamics of these solutions were studied in refs. \cite{Chamblin:1999tk,Chamblin:1999hg}. There, a rich phase structure was uncovered, a  van der Waals--like nature was revealed, including a second order critical point. Such structures can be embedded into an extended thermodynamics \cite{Kastor:2009wy}, where  dynamical~$\Lambda$ defines a pressure $p=-\Lambda/8\pi$ (and its conjugate volume~$V$). In this context, ref. \cite{Kubiznak:2012wp} clarified the resemblance to van der Waals and showed that the system (for $D=4$) has the same universal behaviour near the critical point as the van der Waals gas \footnote{Ref. \cite{Johnson:2013dka} studied the entanglement entropy of the dual theory in the neighbourhood of the second order critical point.}.

The standard semi--classical quantum gravity procedures  \cite{Bekenstein:1973ur,Bekenstein:1974ax,Hawking:1974sw,Hawking:1976de,Gibbons:1976ue}  yield  a temperature $T=Y^\prime(r_+)/4\pi$ for each black hole solution, which depends on $r_+$,~$q$, and~$l$:
\begin{equation}
\label{eq:temperature}
T=\frac{1}{4\pi}\left(\frac{D-3}{r_+}-\frac{q^2(D-3)}{r_+^{2D-5}}+(D-1)\frac{r_+}{l^2}\right)\ ,
\end{equation}
and the entropy is  given by a quarter of the area of the horizon: $S=\omega_{D-2}r_+^{D-2}/4$. The extended thermodynamics \cite{Kastor:2009wy} has all appearances of the length scale~$l$  replaced by the pressure $p$ using the relation 
\begin{equation}
\label{eq:pressure}
p=\frac{(D-1)(D-2)}{16\pi l^2}\ , 
\end{equation}
and the thermodynamic volume turns out to be the flat space volume of the ball bounded by $S^{D-2}$: $V=\omega_{D-2}r_+^{D-1}/(D-1)$.  
So all occurrences of $r_+$ can be traded in for either  an $S$ or a~$V$, as they are not independent.  
In this way, the temperature equation~(\ref{eq:temperature}) defines an equation of state $p(V,T)$. Also, the equation $Y(r_+)=0$ can be rearranged to give an equation for~$m$, which yields the mass $M$ {\it via}  equation~(\ref{eq:mass}). Written in terms of the entropy and pressure $M$ is actually the enthalpy $H(S,p)$ of the system in terms of its natural variables\cite{Kastor:2009wy,Dolan:2011xt}. So we have:
\begin{eqnarray}
\label{eq:equationofstate}
p&=&\frac{(D-2)}{16\pi}\left(\frac{4T}{r_+}-\frac{(D-3)}{r_+^2}+\frac{q^2(D-3)}{r_+^{2D-4}}\right)\ , \nonumber \\
m&=&r_+^{D-3}+\frac{q^2}{r_+^{D-3}}+\frac{16\pi p\, r_+^{D-1}}{(D-1)(D-2)} \ .
\label{eq:enthalpy}
\end{eqnarray}

The main focus here will be the second order critical point, discovered in refs. \cite{Chamblin:1999tk,Chamblin:1999hg}. It is inherited by the extended thermodynamics, and lies on the $p(V,T)$ curve that has a point of inflection: $\partial p/\partial V=\partial^2 p /\partial V^2=0$. The derivatives can also be taken with respect to $r_+$ here, and  the criticql point is defined by the  resulting  values \cite{Kubiznak:2012wp,Gunasekaran:2012dq}:
\begin{eqnarray}
\label{eq:critical}
r_{\rm cr}&=&\{ q^2(D-2)(2D-5)\}^{1/(2D-6)}\nonumber\\
T_{\rm cr}&=&\frac{(D-3)^2}{\pi r_{\rm cr}(2D-5)}\ , \quad p_{\rm cr}=\frac{(D-3)^2}{16\pi r_{\rm cr}^2}\ .
\end{eqnarray}
The case of $D{=}4$ is special. It begins with the observation that $q$ is a length scale in this case, and the precise relations are:
\begin{equation}
\label{eq:critical4}
r_{\rm cr}=\sqrt{6}q\ ,\quad T_{\rm cr}= \frac{1}{3\sqrt{6}\pi q}\ ,  \quad p_{\rm cr}=\frac{1}{96\pi q^2}\ , 
\end{equation}
(with the latter implying a constant scalar force $F=pA=1/4$), and  the resulting entropy and thermodynamic volume are
 $S_{\rm cr}=6\pi q^2$  and $V_{\rm cr}= 8\sqrt{6}\pi q^3.$

In ref. \cite{Johnson:2017hxu}, certain heat engine cycles  in the $(p,V)$ plane that were located in the neighbourhood of this $D=4$ critical point were found to have the interesting property that their efficiency  asymptotes to the Carnot efficiency in the limit of large $q$.   This large $q$ limit is not a high temperature or ideal gas limit where the working substance simplifies in some straightforward sense. Rather, $q$ is  more like a measure of the number of interacting subsystems, as is suggested by the fact that the thermodynamic volume grows with $q$, {\it i.e., } $V\sim q^{(D-1)/(D-3)}$. For $D=4$, since $q$ has dimensions of length, $V\sim q^3$, and at  large $q$ (in this critical neighbourhood),  the subsystems  conspire to provide (for the given choice of cycle) the most ideal trade--off between work done and heat intake allowed by the second law of thermodynamics. 

The idea of a  coupled system being driven to Carnot efficiency at criticality was modelled in the  recent statistical mechanics literature\cite{2015PhRvL.114e0601P,2016NatCo...711895C}, and as pointed out in ref.\cite{Johnson:2017hxu},  this gravitational setting seems to be an   exactly solvable realization of this type of arrangement.

All of this motivates a closer look at the critical region itself, for general $D$. The central object there is the charged black hole with the critical quantities given in equation~(\ref{eq:critical}). The critical black hole mass  parameter $m_{\rm cr}$ and critical cosmological scale $l_{\rm cr}$ are easily computed (using equations~(\ref{eq:enthalpy}), (\ref{eq:critical}) and~(\ref{eq:pressure})), 
and so the metric function  for our critical black hole is:
\begin{eqnarray}
Y_{\rm cr}(r)&=& 1-\frac{m_{\rm cr}}{r^{D-3}}+\frac{q^2}{r^{2D-6}}+\frac{r^2}{l^2_{\rm cr}}\ ,\\
 l_{\rm cr}^2&=&\frac{(D-1)(D-2)}{(D-3)^2}r^2_{\rm cr}\ , \quad m_{\rm cr}=m(r_{\rm cr})\ , \nonumber
\end{eqnarray}
with $l_{\rm cr}^2=36 q^2$ and $M_{\rm cr}=m_{\rm cr}/2=4/\sqrt{6} q$ in the case of $D=4$.

We can further explore the picture of $q$ interacting constituent objects by studying the motion of a point particle of  mass $\mu$  and charge $e$ moving in the background of the critical  charged black hole, in a probe approximation. It  is studied using standard techniques (see e.g. ref.\cite{Chandrasekhar:1985kt,Olivares:2011xb}) and for the critical hole the following effective potential determines the nature of the possible orbits:
\begin{equation}
\label{eq:effectivepotential}
V_{\rm eff}= \frac{eq}{r^{D-3}}+\sqrt{Y_{\rm cr}(r)}\sqrt{\mu^2+\frac{L^2}{r^2}}\ ,
\end{equation}
where $L$ is the angular momentum of the particle. Generically, this potential rises at large $r$ due to the cosmological term, and depending upon $e,\mu$ and $L$, it may have a local minimum at some $r_{\rm min}>r_{\rm cr}$. 

The presence of such a minimum for the particle at rest would suggest the possibility that instead of  (or additionally to) an arrangement where all the constituents are gathered together, there could be some of them  forming a shell at $r_{\rm min}$. This would not match well with the motivating scenario and could even represent a channel of instability, as it does in the case (for flat black holes in AdS) of  dual models of superfluidity and superconductivity signalling the formation of the condensate \cite{Gubser:2008px,Hartnoll:2008vx}.

\begin{figure}[h]
\includegraphics[width=3.0in]{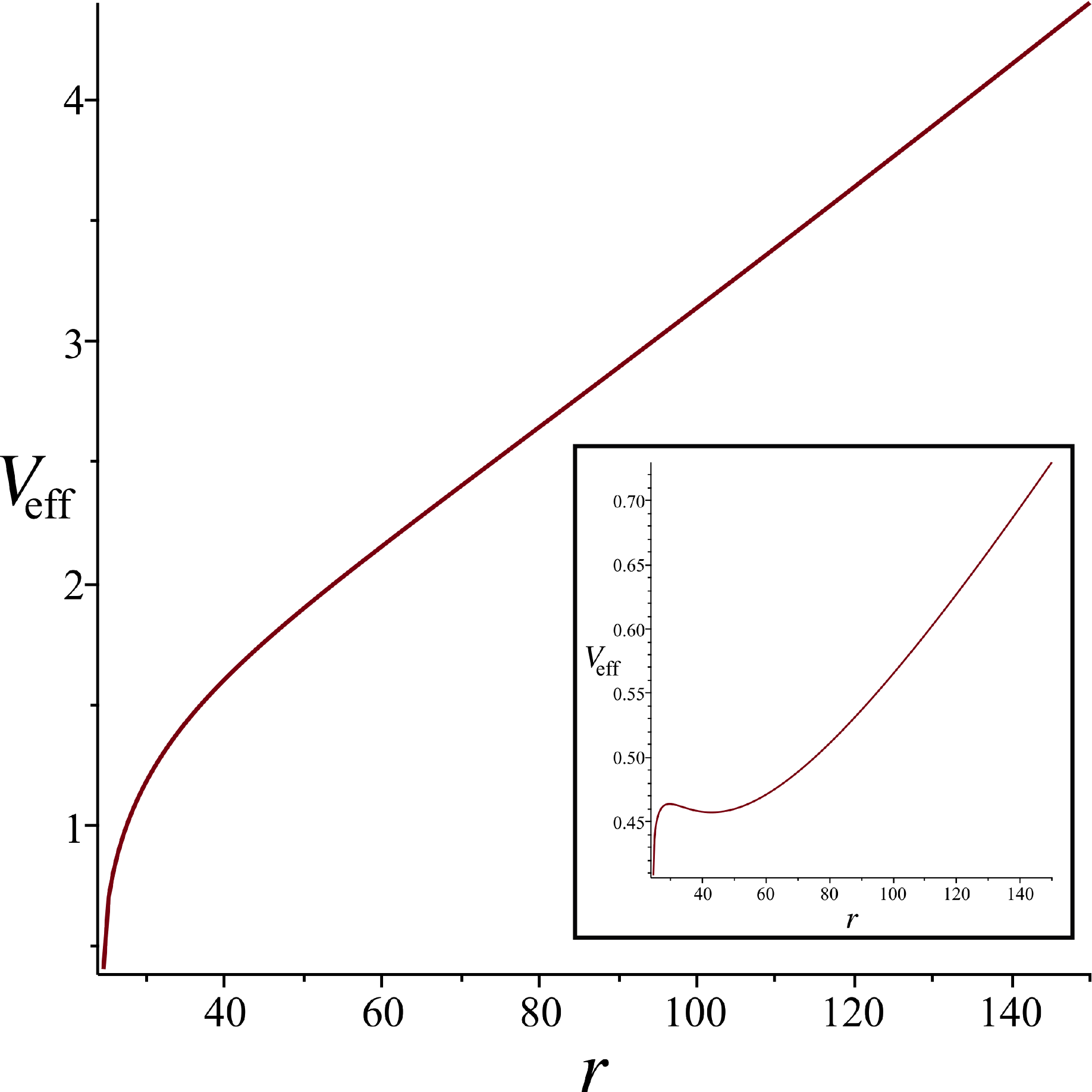} 
   \caption{\footnotesize  Main: The $L=0$ effective potential $V_{\rm eff}(r)$  for $D=4$ when the probe has the critical mass--to--charge ratio $\mu/e=M_{\rm cr}/q=4/\sqrt{6}$, with $q=10$. It is purely attractive toward the black hole. Inset: The $L=0$ $V_{\rm eff}(r)$  in $D=4$ when $\mu/e=0.25$, $q=10$, showing a local minimum.}   \label{fig:effectivepotential}
\end{figure}

Looking at the case of $L=0$, one can seek  a minimum  and find out if it is physical ({\it i.e.} $r_{\rm min}>r_{\rm cr}$) when the probe has the critical charge--to--mass ratio $M_{\rm cr}/q$. This turns out to be difficult to analyze exactly, but some numerical investigation can be  employed. The case of  $D=4$ is presented in figure~\ref{fig:effectivepotential}, where the range of $r$ plotted begins at the critical horizon radius $r_{\rm cr}=\sqrt{6} q$.  For $\mu/e=0.25$, there is a minimum, which disappears after an increase of a few percent. For the critical ratio $\mu/e=M_{\rm cr}/q=4/\sqrt{6}$, there is no minimum.

This result is consistent with the interacting subsystem picture mentioned above. The probe particle with $\mu/e=4/\sqrt{6}$ represents one of the   subsystems while the charge  $q$ critical background is the collective effect of all the others making up the system. There is a purely attractive potential binding all the constituents together. Similar numerical investigations for integer $D$ from 5 to  10 showed analogous behaviour. Indeed, increasing $D$ seems to require increasingly smaller $\mu/e$ in order to generate a minimum.

It is interesting to study the near--horizon geometry of this  solution, while at the same time focussing on the case of large $q$. Recall that the horizon radius $r_+$  is defined by $Y(r=r_+)=0$, and let the neighbourhood of the horizon  be explored by coordinate~$\sigma$. Writing,  $r=r_++\epsilon\sigma$ and $t= \tau/\epsilon$, for small $\epsilon$, the metric becomes (to first order in $\epsilon$):
\begin{eqnarray}
ds^2 &=& -\frac{\sigma Y^\prime}{\epsilon}d\tau^2+\frac{\epsilon}{\sigma Y^\prime}d\sigma^2\\
&&\hskip3cm +\,(r_+^2+2\epsilon\sigma r_+)d\Omega_{D-2}^2\ , \nonumber 
\end{eqnarray}
where $Y^\prime=dY(r)/dr|$ evaluated at $r=r_+$. Now  $Y^\prime= 4\pi T$, and we are interested in the critical hole, so $T$  and $r_+$ are given by $T_{\rm cr}$ and $r_{\rm cr}$ of equation~(\ref{eq:critical}). We  go to the near horizon limit $\epsilon\to0$ while at the same time taking the large $q$ limit by holding  ${\tilde q}^{1/(D-3)}=\epsilon q^{1/(D-3)}$ fixed. The result is:
\begin{equation}
ds^2 = -{(4\pi {\widetilde T}_{\rm cr})\,\,\sigma}d\tau^2+\frac{1}{(4\pi {\widetilde T}_{\rm cr})}\frac{d\sigma^2}{\sigma }+d{\mathbb R}^{D-2},
\end{equation}
where, ${\widetilde T}_{\rm cr}$ is $T_{\rm cr}$ in equation~(\ref{eq:critical}) with $q$ replaced by~${\tilde q}$. Also,  $\Lambda=0$ and $A_t=0$, and  since the $S^{D-2}$ has infinite radius ($r_{\rm cr}$ diverges at large $q$), the metric there is effectively $d{\mathbb R}^{D-2}=dx_1^2+\cdots+ dx_{D-2}^2\ $. 

This result of our double limit is the line element for $D$--dimensional Rindler spacetime. Two things are worth noting here. 
The first is that this is different to the copy of $(1+1)$--dimensional Rindler (times an $S^{D-2}$) that generically appears near the horizon of a non--extremal black hole. The second is that this is a fully decoupled spacetime (in contrast to the usual Rindler$_2\times S^{D-2}$) in the sense that the throat length (the proper distance from the horizon at $r=r_+$ to some finite coordinate distance~$R$), which grows with~$q$, diverges in this limit.  This makes it more analogous to the case of the near--horizon geometry of extremal black holes, which decouples because the throat length diverges due to  the double root of $Y(r)$ in that case (for any $q$). There, the spacetime becomes AdS$_2\times S^{D-2}$. Such decoupled  spacetimes (often with higher dimensional AdS factors) have become the basis for many useful computations for ``dual" strongly coupled systems, as is well known\cite{Maldacena:1997re,Witten:1998qj,Gubser:1998bc,Witten:1998zw}. It would be interesting to explore whether computations starting with the decoupled Rindler$_D$ arising here (for this large $q$ near--horizon limit) might be of use in learning more about the critical systems described by the black holes.

\medskip
 
 \begin{acknowledgments}
CVJ  thanks the  US Department of Energy for support under grant DE--SC0011687,  the Simons Foundation for  a Simons Fellowship (2017), and Amelia for her support and patience.    
\end{acknowledgments}

\bibliographystyle{apsrev4-1}
\bibliography{johnson_critical}

\begin{thebibliography}{28}%
\makeatletter
\providecommand \@ifxundefined [1]{%
 \@ifx{#1\undefined}
}%
\providecommand \@ifnum [1]{%
 \ifnum #1\expandafter \@firstoftwo
 \else \expandafter \@secondoftwo
 \fi
}%
\providecommand \@ifx [1]{%
 \ifx #1\expandafter \@firstoftwo
 \else \expandafter \@secondoftwo
 \fi
}%
\providecommand \natexlab [1]{#1}%
\providecommand \enquote  [1]{``#1''}%
\providecommand \bibnamefont  [1]{#1}%
\providecommand \bibfnamefont [1]{#1}%
\providecommand \citenamefont [1]{#1}%
\providecommand \href@noop [0]{\@secondoftwo}%
\providecommand \href [0]{\begingroup \@sanitize@url \@href}%
\providecommand \@href[1]{\@@startlink{#1}\@@href}%
\providecommand \@@href[1]{\endgroup#1\@@endlink}%
\providecommand \@sanitize@url [0]{\catcode `\\12\catcode `\$12\catcode
  `\&12\catcode `\#12\catcode `\^12\catcode `\_12\catcode `\%12\relax}%
\providecommand \@@startlink[1]{}%
\providecommand \@@endlink[0]{}%
\providecommand \url  [0]{\begingroup\@sanitize@url \@url }%
\providecommand \@url [1]{\endgroup\@href {#1}{\urlprefix }}%
\providecommand \urlprefix  [0]{URL }%
\providecommand \Eprint [0]{\href }%
\providecommand \doibase [0]{http://dx.doi.org/}%
\providecommand \selectlanguage [0]{\@gobble}%
\providecommand \bibinfo  [0]{\@secondoftwo}%
\providecommand \bibfield  [0]{\@secondoftwo}%
\providecommand \translation [1]{[#1]}%
\providecommand \BibitemOpen [0]{}%
\providecommand \bibitemStop [0]{}%
\providecommand \bibitemNoStop [0]{.\EOS\space}%
\providecommand \EOS [0]{\spacefactor3000\relax}%
\providecommand \BibitemShut  [1]{\csname bibitem#1\endcsname}%
\let\auto@bib@innerbib\@empty
\bibitem [{\citenamefont {Bekenstein}(1973)}]{Bekenstein:1973ur}%
  \BibitemOpen
  \bibfield  {author} {\bibinfo {author} {\bibfnamefont {J.~D.}\ \bibnamefont
  {Bekenstein}},\ }\href {\doibase 10.1103/PhysRevD.7.2333} {\bibfield
  {journal} {\bibinfo  {journal} {Phys.Rev.}\ }\textbf {\bibinfo {volume}
  {D7}},\ \bibinfo {pages} {2333} (\bibinfo {year} {1973})}\BibitemShut
  {NoStop}%
\bibitem [{\citenamefont {Bekenstein}(1974)}]{Bekenstein:1974ax}%
  \BibitemOpen
  \bibfield  {author} {\bibinfo {author} {\bibfnamefont {J.~D.}\ \bibnamefont
  {Bekenstein}},\ }\href {\doibase 10.1103/PhysRevD.9.3292} {\bibfield
  {journal} {\bibinfo  {journal} {Phys.Rev.}\ }\textbf {\bibinfo {volume}
  {D9}},\ \bibinfo {pages} {3292} (\bibinfo {year} {1974})}\BibitemShut
  {NoStop}%
\bibitem [{\citenamefont {Hawking}(1975)}]{Hawking:1974sw}%
  \BibitemOpen
  \bibfield  {author} {\bibinfo {author} {\bibfnamefont {S.}~\bibnamefont
  {Hawking}},\ }\href {\doibase 10.1007/BF02345020} {\bibfield  {journal}
  {\bibinfo  {journal} {Commun.Math.Phys.}\ }\textbf {\bibinfo {volume} {43}},\
  \bibinfo {pages} {199} (\bibinfo {year} {1975})}\BibitemShut {NoStop}%
\bibitem [{\citenamefont {Hawking}(1976)}]{Hawking:1976de}%
  \BibitemOpen
  \bibfield  {author} {\bibinfo {author} {\bibfnamefont {S.}~\bibnamefont
  {Hawking}},\ }\href {\doibase 10.1103/PhysRevD.13.191} {\bibfield  {journal}
  {\bibinfo  {journal} {Phys.Rev.}\ }\textbf {\bibinfo {volume} {D13}},\
  \bibinfo {pages} {191} (\bibinfo {year} {1976})}\BibitemShut {NoStop}%
\bibitem [{\citenamefont {Gibbons}\ and\ \citenamefont
  {Hawking}(1977)}]{Gibbons:1976ue}%
  \BibitemOpen
  \bibfield  {author} {\bibinfo {author} {\bibfnamefont {G.~W.}\ \bibnamefont
  {Gibbons}}\ and\ \bibinfo {author} {\bibfnamefont {S.~W.}\ \bibnamefont
  {Hawking}},\ }\href {\doibase 10.1103/PhysRevD.15.2752} {\bibfield  {journal}
  {\bibinfo  {journal} {Phys. Rev.}\ }\textbf {\bibinfo {volume} {D15}},\
  \bibinfo {pages} {2752} (\bibinfo {year} {1977})}\BibitemShut {NoStop}%
\bibitem [{\citenamefont {Kastor}\ \emph {et~al.}(2009)\citenamefont {Kastor},
  \citenamefont {Ray},\ and\ \citenamefont {Traschen}}]{Kastor:2009wy}%
  \BibitemOpen
  \bibfield  {author} {\bibinfo {author} {\bibfnamefont {D.}~\bibnamefont
  {Kastor}}, \bibinfo {author} {\bibfnamefont {S.}~\bibnamefont {Ray}}, \ and\
  \bibinfo {author} {\bibfnamefont {J.}~\bibnamefont {Traschen}},\ }\href
  {\doibase 10.1088/0264-9381/26/19/195011} {\bibfield  {journal} {\bibinfo
  {journal} {Class.Quant.Grav.}\ }\textbf {\bibinfo {volume} {26}},\ \bibinfo
  {pages} {195011} (\bibinfo {year} {2009})},\ \Eprint
  {http://arxiv.org/abs/0904.2765} {arXiv:0904.2765 [hep-th]} \BibitemShut
  {NoStop}%
\bibitem [{Note1()}]{Note1}%
  \BibitemOpen
  \bibinfo {note} {See ref. \cite {Kubiznak:2016qmn} for a review of some of
  the results obtained in this context by using the extended thermodynamics
  framework.}\BibitemShut {Stop}%
\bibitem [{\citenamefont {Johnson}(2017)}]{Johnson:2017hxu}%
  \BibitemOpen
  \bibfield  {author} {\bibinfo {author} {\bibfnamefont {C.~V.}\ \bibnamefont
  {Johnson}},\ }\href@noop {} {\  (\bibinfo {year} {2017})},\ \Eprint
  {http://arxiv.org/abs/1703.06119} {arXiv:1703.06119 [hep-th]} \BibitemShut
  {NoStop}%
\bibitem [{\citenamefont {Johnson}(2014{\natexlab{a}})}]{Johnson:2014yja}%
  \BibitemOpen
  \bibfield  {author} {\bibinfo {author} {\bibfnamefont {C.~V.}\ \bibnamefont
  {Johnson}},\ }\href {\doibase 10.1088/0264-9381/31/20/205002} {\bibfield
  {journal} {\bibinfo  {journal} {Class. Quant. Grav.}\ }\textbf {\bibinfo
  {volume} {31}},\ \bibinfo {pages} {205002} (\bibinfo {year}
  {2014}{\natexlab{a}})},\ \Eprint {http://arxiv.org/abs/1404.5982}
  {arXiv:1404.5982 [hep-th]} \BibitemShut {NoStop}%
\bibitem [{Note2()}]{Note2}%
  \BibitemOpen
  \bibinfo {note} {Here, the conventions of refs.\cite
  {Chamblin:1999tk,Chamblin:1999hg} will be used.}\BibitemShut {Stop}%
\bibitem [{\citenamefont {Chamblin}\ \emph
  {et~al.}(1999{\natexlab{a}})\citenamefont {Chamblin}, \citenamefont
  {Emparan}, \citenamefont {Johnson},\ and\ \citenamefont
  {Myers}}]{Chamblin:1999tk}%
  \BibitemOpen
  \bibfield  {author} {\bibinfo {author} {\bibfnamefont {A.}~\bibnamefont
  {Chamblin}}, \bibinfo {author} {\bibfnamefont {R.}~\bibnamefont {Emparan}},
  \bibinfo {author} {\bibfnamefont {C.~V.}\ \bibnamefont {Johnson}}, \ and\
  \bibinfo {author} {\bibfnamefont {R.~C.}\ \bibnamefont {Myers}},\ }\href@noop
  {} {\bibfield  {journal} {\bibinfo  {journal} {Phys. Rev.}\ }\textbf
  {\bibinfo {volume} {D60}},\ \bibinfo {pages} {064018} (\bibinfo {year}
  {1999}{\natexlab{a}})},\ \Eprint {http://arxiv.org/abs/hep-th/9902170}
  {hep-th/9902170} \BibitemShut {NoStop}%
\bibitem [{\citenamefont {Chamblin}\ \emph
  {et~al.}(1999{\natexlab{b}})\citenamefont {Chamblin}, \citenamefont
  {Emparan}, \citenamefont {Johnson},\ and\ \citenamefont
  {Myers}}]{Chamblin:1999hg}%
  \BibitemOpen
  \bibfield  {author} {\bibinfo {author} {\bibfnamefont {A.}~\bibnamefont
  {Chamblin}}, \bibinfo {author} {\bibfnamefont {R.}~\bibnamefont {Emparan}},
  \bibinfo {author} {\bibfnamefont {C.~V.}\ \bibnamefont {Johnson}}, \ and\
  \bibinfo {author} {\bibfnamefont {R.~C.}\ \bibnamefont {Myers}},\ }\href@noop
  {} {\bibfield  {journal} {\bibinfo  {journal} {Phys. Rev.}\ }\textbf
  {\bibinfo {volume} {D60}},\ \bibinfo {pages} {104026} (\bibinfo {year}
  {1999}{\natexlab{b}})},\ \Eprint {http://arxiv.org/abs/hep-th/9904197}
  {hep-th/9904197} \BibitemShut {NoStop}%
\bibitem [{\citenamefont {Kubiznak}\ and\ \citenamefont
  {Mann}(2012)}]{Kubiznak:2012wp}%
  \BibitemOpen
  \bibfield  {author} {\bibinfo {author} {\bibfnamefont {D.}~\bibnamefont
  {Kubiznak}}\ and\ \bibinfo {author} {\bibfnamefont {R.~B.}\ \bibnamefont
  {Mann}},\ }\href {\doibase 10.1007/JHEP07(2012)033} {\bibfield  {journal}
  {\bibinfo  {journal} {JHEP}\ }\textbf {\bibinfo {volume} {1207}},\ \bibinfo
  {pages} {033} (\bibinfo {year} {2012})},\ \Eprint
  {http://arxiv.org/abs/1205.0559} {arXiv:1205.0559 [hep-th]} \BibitemShut
  {NoStop}%
\bibitem [{Note3()}]{Note3}%
  \BibitemOpen
  \bibinfo {note} {Ref. \cite {Johnson:2013dka} studied the entanglement
  entropy of the dual theory in the neighbourhood of the second order critical
  point.}\BibitemShut {Stop}%
\bibitem [{\citenamefont {Dolan}(2011)}]{Dolan:2011xt}%
  \BibitemOpen
  \bibfield  {author} {\bibinfo {author} {\bibfnamefont {B.~P.}\ \bibnamefont
  {Dolan}},\ }\href {\doibase 10.1088/0264-9381/28/23/235017} {\bibfield
  {journal} {\bibinfo  {journal} {Class.Quant.Grav.}\ }\textbf {\bibinfo
  {volume} {28}},\ \bibinfo {pages} {235017} (\bibinfo {year} {2011})},\
  \Eprint {http://arxiv.org/abs/1106.6260} {arXiv:1106.6260 [gr-qc]}
  \BibitemShut {NoStop}%
\bibitem [{\citenamefont {Gunasekaran}\ \emph {et~al.}(2012)\citenamefont
  {Gunasekaran}, \citenamefont {Mann},\ and\ \citenamefont
  {Kubiznak}}]{Gunasekaran:2012dq}%
  \BibitemOpen
  \bibfield  {author} {\bibinfo {author} {\bibfnamefont {S.}~\bibnamefont
  {Gunasekaran}}, \bibinfo {author} {\bibfnamefont {R.~B.}\ \bibnamefont
  {Mann}}, \ and\ \bibinfo {author} {\bibfnamefont {D.}~\bibnamefont
  {Kubiznak}},\ }\href {\doibase 10.1007/JHEP11(2012)110} {\bibfield  {journal}
  {\bibinfo  {journal} {JHEP}\ }\textbf {\bibinfo {volume} {1211}},\ \bibinfo
  {pages} {110} (\bibinfo {year} {2012})},\ \Eprint
  {http://arxiv.org/abs/1208.6251} {arXiv:1208.6251 [hep-th]} \BibitemShut
  {NoStop}%
\bibitem [{\citenamefont {{Polettini}}\ \emph {et~al.}(2015)\citenamefont
  {{Polettini}}, \citenamefont {{Verley}},\ and\ \citenamefont
  {{Esposito}}}]{2015PhRvL.114e0601P}%
  \BibitemOpen
  \bibfield  {author} {\bibinfo {author} {\bibfnamefont {M.}~\bibnamefont
  {{Polettini}}}, \bibinfo {author} {\bibfnamefont {G.}~\bibnamefont
  {{Verley}}}, \ and\ \bibinfo {author} {\bibfnamefont {M.}~\bibnamefont
  {{Esposito}}},\ }\href {\doibase 10.1103/PhysRevLett.114.050601} {\bibfield
  {journal} {\bibinfo  {journal} {Phys. Rev. Lett.}\ }\textbf {\bibinfo
  {volume} {114}},\ \bibinfo {eid} {050601} (\bibinfo {year} {2015})},\ \Eprint
  {http://arxiv.org/abs/1409.4716} {arXiv:1409.4716 [cond-mat.stat-mech]}
  \BibitemShut {NoStop}%
\bibitem [{\citenamefont {{Campisi}}\ and\ \citenamefont
  {{Fazio}}(2016)}]{2016NatCo...711895C}%
  \BibitemOpen
  \bibfield  {author} {\bibinfo {author} {\bibfnamefont {M.}~\bibnamefont
  {{Campisi}}}\ and\ \bibinfo {author} {\bibfnamefont {R.}~\bibnamefont
  {{Fazio}}},\ }\href {\doibase 10.1038/ncomms11895} {\bibfield  {journal}
  {\bibinfo  {journal} {Nat. Commun.}\ }\textbf {\bibinfo {volume} {7}},\
  \bibinfo {eid} {11895} (\bibinfo {year} {2016})},\ \Eprint
  {http://arxiv.org/abs/1603.05024} {arXiv:1603.05024 [cond-mat.stat-mech]}
  \BibitemShut {NoStop}%
\bibitem [{\citenamefont {Chandrasekhar}(1985)}]{Chandrasekhar:1985kt}%
  \BibitemOpen
  \bibfield  {author} {\bibinfo {author} {\bibfnamefont {S.}~\bibnamefont
  {Chandrasekhar}},\ }\href {\doibase 10.1007/978-94-009-6469-3_2} {\emph
  {\bibinfo {title} {{The Mathematical Theory of Black Holes, }}}}\ (\bibinfo
  {publisher} {Oxford, UK: Clarendon},\ \bibinfo {year} {1985})\BibitemShut
  {NoStop}%
\bibitem [{\citenamefont {Olivares}\ \emph {et~al.}(2011)\citenamefont
  {Olivares}, \citenamefont {Saavedra}, \citenamefont {Villanueva},\ and\
  \citenamefont {Leiva}}]{Olivares:2011xb}%
  \BibitemOpen
  \bibfield  {author} {\bibinfo {author} {\bibfnamefont {M.}~\bibnamefont
  {Olivares}}, \bibinfo {author} {\bibfnamefont {J.}~\bibnamefont {Saavedra}},
  \bibinfo {author} {\bibfnamefont {J.~R.}\ \bibnamefont {Villanueva}}, \ and\
  \bibinfo {author} {\bibfnamefont {C.}~\bibnamefont {Leiva}},\ }\href
  {\doibase 10.1142/S0217732311037261} {\bibfield  {journal} {\bibinfo
  {journal} {Mod. Phys. Lett.}\ }\textbf {\bibinfo {volume} {A26}},\ \bibinfo
  {pages} {2923} (\bibinfo {year} {2011})},\ \Eprint
  {http://arxiv.org/abs/1101.0748} {arXiv:1101.0748 [gr-qc]} \BibitemShut
  {NoStop}%
\bibitem [{\citenamefont {Gubser}(2008)}]{Gubser:2008px}%
  \BibitemOpen
  \bibfield  {author} {\bibinfo {author} {\bibfnamefont {S.~S.}\ \bibnamefont
  {Gubser}},\ }\href {\doibase 10.1103/PhysRevD.78.065034} {\bibfield
  {journal} {\bibinfo  {journal} {Phys.Rev.}\ }\textbf {\bibinfo {volume}
  {D78}},\ \bibinfo {pages} {065034} (\bibinfo {year} {2008})},\ \Eprint
  {http://arxiv.org/abs/0801.2977} {arXiv:0801.2977 [hep-th]} \BibitemShut
  {NoStop}%
\bibitem [{\citenamefont {Hartnoll}\ \emph {et~al.}(2008)\citenamefont
  {Hartnoll}, \citenamefont {Herzog},\ and\ \citenamefont
  {Horowitz}}]{Hartnoll:2008vx}%
  \BibitemOpen
  \bibfield  {author} {\bibinfo {author} {\bibfnamefont {S.~A.}\ \bibnamefont
  {Hartnoll}}, \bibinfo {author} {\bibfnamefont {C.~P.}\ \bibnamefont
  {Herzog}}, \ and\ \bibinfo {author} {\bibfnamefont {G.~T.}\ \bibnamefont
  {Horowitz}},\ }\href {\doibase 10.1103/PhysRevLett.101.031601} {\bibfield
  {journal} {\bibinfo  {journal} {Phys. Rev. Lett.}\ }\textbf {\bibinfo
  {volume} {101}},\ \bibinfo {pages} {031601} (\bibinfo {year} {2008})},\
  \Eprint {http://arxiv.org/abs/0803.3295} {arXiv:0803.3295 [hep-th]}
  \BibitemShut {NoStop}%
\bibitem [{\citenamefont {Maldacena}(1998)}]{Maldacena:1997re}%
  \BibitemOpen
  \bibfield  {author} {\bibinfo {author} {\bibfnamefont {J.~M.}\ \bibnamefont
  {Maldacena}},\ }\href@noop {} {\bibfield  {journal} {\bibinfo  {journal}
  {Adv. Theor. Math. Phys.}\ }\textbf {\bibinfo {volume} {2}},\ \bibinfo
  {pages} {231} (\bibinfo {year} {1998})},\ \Eprint
  {http://arxiv.org/abs/hep-th/9711200} {hep-th/9711200} \BibitemShut {NoStop}%
\bibitem [{\citenamefont {Witten}(1998{\natexlab{a}})}]{Witten:1998qj}%
  \BibitemOpen
  \bibfield  {author} {\bibinfo {author} {\bibfnamefont {E.}~\bibnamefont
  {Witten}},\ }\href@noop {} {\bibfield  {journal} {\bibinfo  {journal} {Adv.
  Theor. Math. Phys.}\ }\textbf {\bibinfo {volume} {2}},\ \bibinfo {pages}
  {253} (\bibinfo {year} {1998}{\natexlab{a}})},\ \Eprint
  {http://arxiv.org/abs/hep-th/9802150} {hep-th/9802150} \BibitemShut {NoStop}%
\bibitem [{\citenamefont {Gubser}\ \emph {et~al.}(1998)\citenamefont {Gubser},
  \citenamefont {Klebanov},\ and\ \citenamefont {Polyakov}}]{Gubser:1998bc}%
  \BibitemOpen
  \bibfield  {author} {\bibinfo {author} {\bibfnamefont {S.~S.}\ \bibnamefont
  {Gubser}}, \bibinfo {author} {\bibfnamefont {I.~R.}\ \bibnamefont
  {Klebanov}}, \ and\ \bibinfo {author} {\bibfnamefont {A.~M.}\ \bibnamefont
  {Polyakov}},\ }\href@noop {} {\bibfield  {journal} {\bibinfo  {journal}
  {Phys. Lett.}\ }\textbf {\bibinfo {volume} {B428}},\ \bibinfo {pages} {105}
  (\bibinfo {year} {1998})},\ \Eprint {http://arxiv.org/abs/hep-th/9802109}
  {hep-th/9802109} \BibitemShut {NoStop}%
\bibitem [{\citenamefont {Witten}(1998{\natexlab{b}})}]{Witten:1998zw}%
  \BibitemOpen
  \bibfield  {author} {\bibinfo {author} {\bibfnamefont {E.}~\bibnamefont
  {Witten}},\ }\href@noop {} {\bibfield  {journal} {\bibinfo  {journal} {Adv.
  Theor. Math. Phys.}\ }\textbf {\bibinfo {volume} {2}},\ \bibinfo {pages}
  {505} (\bibinfo {year} {1998}{\natexlab{b}})},\ \Eprint
  {http://arxiv.org/abs/hep-th/9803131} {hep-th/9803131} \BibitemShut {NoStop}%
\bibitem [{\citenamefont {Kubiznak}\ \emph {et~al.}(2017)\citenamefont
  {Kubiznak}, \citenamefont {Mann},\ and\ \citenamefont
  {Teo}}]{Kubiznak:2016qmn}%
  \BibitemOpen
  \bibfield  {author} {\bibinfo {author} {\bibfnamefont {D.}~\bibnamefont
  {Kubiznak}}, \bibinfo {author} {\bibfnamefont {R.~B.}\ \bibnamefont {Mann}},
  \ and\ \bibinfo {author} {\bibfnamefont {M.}~\bibnamefont {Teo}},\ }\href
  {\doibase 10.1088/1361-6382/aa5c69} {\bibfield  {journal} {\bibinfo
  {journal} {Class. Quant. Grav.}\ }\textbf {\bibinfo {volume} {34}},\ \bibinfo
  {pages} {063001} (\bibinfo {year} {2017})},\ \Eprint
  {http://arxiv.org/abs/1608.06147} {arXiv:1608.06147 [hep-th]} \BibitemShut
  {NoStop}%
\bibitem [{\citenamefont {Johnson}(2014{\natexlab{b}})}]{Johnson:2013dka}%
  \BibitemOpen
  \bibfield  {author} {\bibinfo {author} {\bibfnamefont {C.~V.}\ \bibnamefont
  {Johnson}},\ }\href {\doibase 10.1007/JHEP03(2014)047} {\bibfield  {journal}
  {\bibinfo  {journal} {JHEP}\ }\textbf {\bibinfo {volume} {03}},\ \bibinfo
  {pages} {047} (\bibinfo {year} {2014}{\natexlab{b}})},\ \Eprint
  {http://arxiv.org/abs/1306.4955} {arXiv:1306.4955 [hep-th]} \BibitemShut
  {NoStop}%
\end{thebibliography}%

\end{document}